\documentstyle[12pt]{article}
\begin{document}
\newcommand{\scr}{\sin^2 \hat{\theta}_W (m_Z)}
\newcommand{\slept}{\sin^2 \theta_{eff}^{lept}}
\newcommand{\smallw}{{\scriptscriptstyle W}}
\newcommand{\smallz}{{\scriptscriptstyle Z}}
\newcommand{\sef}{\sin^2 \theta_\smallw^{eff}}
\newcommand{\msbar}{\rm{\overline{MS}}}
\newcommand{\be}{\begin{eqnarray}}
\newcommand{\en}{\end{eqnarray}}
\newcommand{\kcar}{\hat{k}}
\newcommand{\mc}{\hat{m}}
\newcommand{\mt}{\hat{m_t}(M_t)}
\newcommand{\scar}{\hat{s}}
\newcommand{\cc}{\hat{c}}
\newcommand{\alc}{\hat{\alpha}}
\newcommand{\rhoc}{\hat{\rho}}
\newcommand{\as}{\alpha_s}
\newcommand{\nn}{\noindent}
\newcommand{\ew}{electroweak~}
\newcommand{\mz}{m_\smallz}
\newcommand{\mw}{m_\smallw}
\newcommand{\dr}{\Delta\rho}
\newcommand{\drf}{(\Delta\rho)_f}
\newcommand{\dq}{\delta_{\scriptscriptstyle QCD}}
\newcommand{\Dq}{\Delta_{\scriptscriptstyle QCD}}
\newcommand{\xiq}{\xi_{\scriptscriptstyle Q}}
\newcommand{\Eq}[1]{Eq.(\ref{#1})}
\newcommand{\Ref}[1]{Ref.\cite{#1}}
\newcommand{\Or}{{\cal O}}
\newcommand{\non}{\nonumber}
\newcommand{\PL}{Phys. Lett.\ }
\newcommand{\NP}{Nucl. Phys.\ }
\newcommand{\PR}{Phys. Rev.\ }

\pagestyle{empty}
{\flushright{ NYU--TH--95/05/02\\
 hep-ph/9505426\\
 May 1995\\}}

\vfill
\begin{center}
\large{\bf Renormalon Contributions to $\dr$}
\end{center}

\vspace*{.8cm}

\centerline{\sc Paolo Gambino
 and Alberto Sirlin}

\vspace*{.4cm}

\centerline{ Department of Physics, New York University, 4 Washington
Place,}
\centerline{ New York, NY 10003, USA\footnote{
E-mail: gambino@acf2.nyu.edu;
sirlin@mafalda.physics.nyu.edu.}.}
\vfill

\begin{center}
\parbox{14cm}
{\begin{center}
ABSTRACT \end{center}
\vspace*{0.15cm}
Leading QCD vacuum polarization contributions
to the \ew parameter $\dr$ are evaluated numerically
using several different prescriptions
for the gluon self-energy. Simple theoretical estimates of the asymptotic
behavior are given. The results show
a significant contribution from the leading infrared renormalons
when $\dr$  is expressed  in terms  of the top-quark pole
mass and its absence  when the  $\msbar$ running mass is employed.
The calculations are applied to estimate higher order QCD contributions
to $\dr$.
}
\end{center}
\vfill
\newpage\pagestyle{plain}
\pagenumbering{arabic}
\nn The higher order effects arising
in the ratio $M/\mc(M)$ ($M$ is the pole mass of
a quark and $\mc$ the $\msbar$ running mass)
from a single chain of vacuum polarization diagrams
has been recently studied
in considerable detail \cite{ps,be}.
In particular, in Ref.\,\cite{ps} it has been emphasized that the basic
building block  in these calculations is a gauge-dependent
amplitude, and therefore there is a considerable latitude in its definition.
This can  be explicitly illustrated by considering the one-loop gluon
vacuum polarization diagram evaluated in the $\xiq$ Background-Field-Gauge
(BFG) \cite{BFG}:
\be
\left.\Pi(k^2,\xiq)\right|^{\msbar}_{BFG}=
\frac{\as(\mu)}{4\pi}\,b\left\{\ln\left(\frac{-\mu^2}{k^2}\right) + \frac{5}{3}
+\frac4{b} \left[ 1 - \frac3{16} (1-\xiq)(7+\xiq)\right]\right\},
\label{vpf}
\en
where $b=11- 2 n_f/3$ is the coefficient of the leading term in the QCD
$\beta$-function, $n_f$ the number of massless fermions, $k$ the external
gluon momentum, $\xiq$ is the gauge parameter associated with the quantum
loops, and the superscript $\msbar$ means that the $\msbar$ renormalization has
been carried out.
An important feature of \Eq{vpf} is that it includes correctly the logarithms
associated with the running of $\as$.
However, the accompanying constant depends on the choice of $\xiq$.
Most current evaluations of QCD vacuum polarization effects are  based
on the leading $n_f$ approximation, coupled with a non-abelianization
prescription \cite{be}: starting from the complete fermionic contribution to
the gluon self-energy, one includes the dominant gluonic component
by replacing $-2n_f /3\to b$. This is equivalent to the neglect of the
$1/b$ term in the expression between curly brackets  in \Eq{vpf}. For brevity,
we will refer to this procedure as the ``non-abelianization'' (NA)
approach. In \Ref{ps} several different prescriptions for
 the basic building block
were discussed.
In this paper we extend these considerations to evaluate the higher-order
effects induced by a chain
of vacuum polarization graphs in the QCD corrections
to $\drf$, the fermionic component of $\dr$. We recall that
this amplitude plays a significant role in \ew physics, as it contains
the leading contributions, for large $M_t$, of the basic corrections
$\Delta r$, $\Delta \hat{r}$, and $\Delta\hat{\rho}$ \cite{sir80,msbar}.

A very convenient framework to carry out this analysis is provided by the
work of Smith and Voloshin (S-V) \cite{voloshin}.
 Setting $M_b=0$ and writing
\be
\drf= \frac{3G_\mu}{8\sqrt{2} \pi^2} \, M_t^2 \,(1+\dq),
\label{2}
\en
where $M_t$ is the top-quark pole  mass and $\dq$ is the QCD correction,
these authors have shown that the one-loop vacuum polarization insertion in the
 $\Or(\as)$ correction is given  by
\be
 (\dq)_{v.p.}= \frac2{3} \frac{\as(\mu)}{\pi}\,\int_0^\infty
[w(x)+s(x)] (1+\Pi(-\kappa^2) ) \,dx^2,
\label{int}
\en
where $x=\kappa/\mu$, $\kappa^2=-k^2$ is the  invariant euclidean gluon
 momentum, and the weight functions $w(x)$ and $s(x)$ are given  by
\be
w(x)&=& 2\,\frac{36+70x^2+48 x^4 + 12 x^6 + x^8}{x(4+x^2)^{3/2}}
\tanh^{-1}\left(x\over\sqrt{4+x^2}\right)+ x^4 \ln x^2\non\\
&&-2 \,\frac{(1+x^2)^3}{x^2} \ln (1+x^2) - 2 \,\frac{5+2x^2}{4+x^2},
\en
\be
s(x)= \frac{x^4+2x^2 -8}{2x\sqrt{4+x^2}} - \frac{x^2}{2}.
\en
The function $w(x)$ occurs in the combination of vacuum polarization
contributions to $\drf$ involving the $(t,b)$ isodoublet, while $s(x)$ is
associated with the mass counterterm corresponding to the $M_t$ pole
definition.
Defining the leading vacuum polarization effects as the contributions of $\Or
(\as^n)$ involving $n-1$ one-loop vacuum polarization insertions
(i.e. the maximum number of insertions),
it is clear that an analogous approach can be employed in their evaluation.
As the integral in \Eq{int} is convergent, it suffices to replace
$1+\Pi(-\kappa^2) \to (1-\Pi(-\kappa^2))^{-1}$ with
 $n_f=5$ (i.e. $b=23/3$), expand the geometric series and carry out
the integrations numerically.

Using the generic expression:
\be
\dq= - \frac{8 a(M_t)}{3} \ \sum_{n=0}^\infty
\rho_n \left(\frac{b\,a(M_t)}{2}\right)^n n!,
\label{dq}
\en
where $a\equiv\as/\pi$, the coefficients $\rho_n $ are displayed
in Table 1 up to $n=10$ for five different prescriptions
to evaluate the basic building block $\Pi(k^2)$: 1) NA, in which case
the $\rho_n$ coefficients are denoted as $a_n$; ii) Pinch Technique (PT)
\cite{PT} which leads to the same expression as  the  $\xiq=1 $ BFG
($\rho_n\to p_n$); iii) $\xiq=0$ (or Landau BFG) with $\rho_n\to\ell_n$;
iv) $\xiq=-3$ BFG, which corresponds to the minimum of the non-logarithmic
contribution in \Eq{vpf} ($\rho_n\to v_n$); v) $\msbar$ running coupling
$\as(\kappa)$  ($\rho_n\to r_n$).
It was pointed out in \Ref{ps} that, by a rather remarkable coincidence,
the $\xiq=-3$ self-energy equals the $\Or(\as)$ correction to the QCD
potential between two infinitely heavy quarks \cite{hq}, and it is thus
amenable to a physical interpretation.
We have also included the elementary running-coupling
prescription, in which only the logarithmic term in \Eq{vpf} is retained.
Although it cannot be obtained from \Eq{vpf} for any real value of $\xiq$,
this approach is perhaps the most intuitively simple,
and it is frequently employed in the literature \cite{bigi,mueller}.
Another attractive feature is that $\as(\kappa)$ is gauge-invariant.

The evaluation procedure involves a numerical integration up to a
cutoff $\Lambda$, with the tail beyond the cutoff evaluated analytically
using the asymptotic expansion for the weight functions $w(x)$ and $s(x)$.
Different values of $\Lambda$ were employed to check the stability of the
evaluation, and a similar method was employed for the very small $x$
integration region. Typically, the variation found is of the order
of a few parts in $10^8$. An important consistency check involves the $a_1$
coefficient, as this must lead exactly to the $\Or(\as^2(M_t))$ contribution
proportional to $n_f$. Replacing $b\to - 2n_f/3$ in the $n=1$ term of \Eq{dq},
we have $(8/9) a_1 n_f a^2(M_t)$. Keeping eight
significant figures our numerical evaluation gives $a_1= 2.0094894$,
which leads to 1.7862128 $n_f a^2(M_t)$. To this accuracy, this value coincides
with the very precise determination reported in \Ref{3loopa}.
Other consistency checks involve the predicted values of the asymptotic
coefficients, which will be discussed later.

A glance at Table 1 shows that all the coefficients are positive and the
various sequences converge to well-defined limits, with the asymptotic behavior
setting in remarkably quickly. This is analogous to the result for $M_t/\mt$,
obtained by analytic methods \cite{ps,be}. The structure of \Eq{dq}, involving
the factorial enhancement, the asymptotically constant $\rho_n$ of fixed sign,
 and the expansion variable $ba(M_t)/2$, shows the characteristic Borel
non-summable behavior associated with the leading infrared-renormalon
contribution. In our case, this is the pole in the Borel plane at $u\equiv
bt/(4\pi)=1/2$, where $t$ is the Borel parameter \cite{ps,be}. Physically,
this behavior emerges because in the evaluation of Feynman amplitudes the gluon
self-energy (or alternatively the strong interaction coupling) becomes
very large at low values of $k/\mu$, and perturbation theory breaks down
\cite{mueller}.

Combining the above results with those obtained in \Ref{ps}, one can readily
find the corresponding expansions for $\Dq$, the QCD correction when
$\dr$ is expressed in terms of $\mc_t^2(M_t)$:
\be
\drf= \frac{3G_\mu}{8\sqrt{2} \pi^2} \, \mc_t^2(M_t) (1+\Dq).
\label{msbardr}
\en
Defining again the leading vacuum polarization effects as the contributions
arising, in a given order in $\as$, from the maximum number of vacuum
polarization insertions, we have
\be
\left(\frac{M_t}{\mc_t^2(M_t)}\right)^2= 1
+ \frac8{3} a(M_t)\sum_{n=0}^\infty \mu_n \left(\frac{b\,a(M_t)}{2}\right)^n
n!,
\label{ratio}
\en
where the $\mu_n$ coefficients are evaluated, for the i)-iv) prescriptions,
in \Ref{ps}.
We have also calculated the coefficients
$\mu_n$ for the running coupling approach, by the same analytic
methods. Combining Eqs.(\ref{vpf}) and (\ref{dq}-\ref{ratio}), one obtains
for the leading contributions
\be
\Dq= - \frac8{3} a(M_t) \,
 \sum_{n=0}^\infty \hat{\rho}_n
\left(\frac{ba(M_t)}{2}\right)^n n! ,
\label{Dq}
\en
with $\hat{\rho}_n= \rho_n -  \mu_n$.
Using the notation
$\hat{\rho}_n\to \hat{p}_n,\hat{a}_n,\hat{\ell}_n,\hat{v}_n, \hat{r}_n$
for the PT, NA, $\xiq=0$, $\xiq=-3$, and running coupling prescriptions,
respectively, these coefficients are displayed in Table 2 up to $n=10$.
Comparing Tables 1 and 2, we see that the pattern of the
$M_t^2$ and $\mc_t^2(M_t)$ expansions is dramatically different.
In the latter case, the coefficients are much  smaller, decrease as $n$
becomes large and the signs alternate after $n=2$.

The general features and many of the detailed results displayed in Tables
1 and 2 can be understood on the basis of  very simple asymptotic estimates.
Expressing the gluon self-energy in the abbreviated form
$\Pi(-\kappa^2)= (a(M_t)b/4) \ln(1/\zeta^2 x^2)$, where $\zeta^2=e^{-P^2}$
and $P^2=5/3 + (4/b) [1-3(1-\xiq)(7+\xiq)/16]$, we have
\be
 (\dq)_{v.p.}= \frac2{3} a(M_t)\,\sum_{n=0}^\infty \left(\frac{a(M_t)b}{2}
\right)^n \int_0^\infty 2x\,
[w(x)+s(x)]  \ln^n \hspace*{-.8mm}\left(\frac1{\zeta x}\right) \,dx.
\label{int2}
\en
As the infrared-renormalon contributions involve the very small $x$ region,
we introduce an ultraviolet cutoff $x_{max}$ and employ the expansion
$2x[w(x)+s(x)]= -4 +\frac3{2} x^2 + 2 x^3 +\ldots$ ($x\ll1$). Setting
$y=\ln(1/\zeta x)$ as integration variable, and choosing $x_{max}=1/\zeta$
(where $\Pi(-\kappa^2)=0$), the integrals reduce to factorials and we obtain
\be
 (\dq)_{v.p.}^{IR}= \frac2{3} a(M_t)\,\sum_{n=0}^\infty \left(\frac{a(M_t)b}{2}
\right)^n n! \ \frac1{\zeta}\left[ -4 + \frac1{2\zeta^2 3^n }+
\frac1{2\zeta^3 4^n} +\ldots\right].
\label{dqir}
\en
These contributions are associated with poles on the positive real axis
of the Borel plane (infrared renormalons). The first term arises from
the pole nearest to the origin ($u=1/2$) and gives the  leading asymptotic
behavior for large $n$, while the others involve more distant renormalons
($u=3/2,2\ldots$) and represent subleading contributions.
To study the dominant asymptotic contributions  from the
ultraviolet region we introduce an infrared cutoff $x_{min}$ in \Eq{int2}
and use the asymptotic expansion $2x[w(x)+s(x)]= 24(\ln x-1)/x^3 +
(157-192\ln x)/x^5+\ldots$ ($x\gg1$). Choosing $x_{min}=x_{max}=1/\zeta$
and proceeding as before, we obtain
\be
 (\dq)_{v.p.}^{UV}&=& \frac2{3} a(M_t)\,\sum_{n=0}^\infty \left(
\frac{a(M_t)b}{4}\right)^n
(-1)^n \ n! \ 6\,\zeta^2\,\nonumber\\
&\times&\left\{n-1-2\ln\zeta - \frac{\zeta^2}{2^n }
\left[2n- \frac{109}{24} - 8 \ln\zeta\right]
 +\ldots\right\}.
\label{dquv}
\en
The two terms between curly brackets correspond to the leading and
next-to-leading ultraviolet renormalons (poles on the negative real axis
of the Borel plane). The unusual contributions proportional to $n$ within
the curly brackets arise from the additional logarithms from the $w(x)$
expansion, and correspond to singularities of the form $u/(1+u)^2$,
$u/(2+u)^2$ in the Borel
plane (i.e. combinations of simple and double poles).
Eqs.\,(\ref{dq},\ref{dqir},\ref{dquv}) imply $\rho_n\to 1/\zeta=e^{P^2/2}$ as
$n\to\infty$, the same asymptotic behavior as the $\mu_n$ \cite{ps}, so that
$\rhoc_n\to 0$. In fact, it is very instructive to compare \Eq{dqir}
with the corresponding expansion in $M^2/\mc^2(M)$. One readily discovers
that the two leading infrared-renormalon expansions have equal magnitude and
opposite sign and therefore cancel in $\Dq$! \
Combining Eqs.\,(\ref{dqir},\ref{dquv})
with the corresponding expansions for $M^2/\mc^2(M)$, one obtains
\be
(\Dq)_{v.p.}^{UV}&=& \frac2{3} a(M_t)\,\sum_{n=0}^\infty \left(
\frac{a(M_t)b}{4}\right)^n
(-1)^n \ n! \ 6\,\zeta^2\,\nonumber\\
&\times&\left\{n-\frac5{3} -2\ln\zeta - \frac{\zeta^2}{2^n }
\left[2n- \frac{127}{24} - 8 \ln\zeta\right]
 +\ldots\right\}.
\label{Dquv}
\en
\be
 (\Dq)_{v.p.}^{IR}= \frac2{3} a(M_t)\,\sum_{n=0}^\infty \left(\frac{a(M_t)b}{8}
\right)^n n! \ \frac3{4\zeta^4} +\ldots
\label{Dqir}
\en
We see that, in contrast with $\dq$, the leading asymptotic behavior in $\Dq$
is associated with the nearest ultraviolet renormalon ($u=-1$),
which explains the alternating signs  in Table 2 and the much reduced
coefficients relative to Table 1. The simple asymptotic formulae of
Eqs.\,(\ref{dqir},\ref{dquv}) approximate the $\rho_n$
coefficients (Table 1) with considerable accuracy for $n\ge 2$. On the other
 hand, Eqs.\,(\ref{Dquv},\ref{Dqir}) give an accurate estimate of the $\rhoc_n
$ coefficients (Table 2) for $n\ge N$, where $N$ depends on the prescription
employed. We illustrate this for the $v_n$ and $\hat{v}_n$ coefficients
in the last column of Tables 1 and 2. We see that the $v_n$ and
$\hat{v}_n$ fits are  quite
 precise for $n\ge 1 $ and $n\ge 2$, respectively.
 We have extended this comparison up to the 17th
coefficient and find that the numerical and asymptotic evaluations of
$\hat{v}_{17}=9.79\times 10^{-5}$ differ by only 1.6$\times 10^{-8}$.
As the leading asymptotic contribution for $\Dq$ arises from the $u=-1$
ultraviolet renormalon, it is Borel summable and can be expressed in terms
of ordinary functions and exponential integrals.
Subtracting the $n=0,1$ terms from this sum and using the $\xiq=-3$
prescription, we find that, for $M_t=175$GeV, it contributes only 4.3$\times
10^{-4}$ to $\Dq$. Thus, in contrast with the large effects from infrared
renormalons in $\dq$, the $n\ge2$ terms in the leading ultraviolet
renormalon expansion give a small contribution to $\Dq$.

In the limit $M_b=0$, recent exact calculations \cite{3loopa,3loopb}
lead to the expansion
\be
\dq = - 2.8599 \ a(M_t) - 14.594 \ a^2(M_t) + \ldots
\label{12}
\en
Using \Eq{dq} and $p_1$, $a_1$, $\ell_1$, $v_1$, and $r_1$ from Table 1,
the $\Or(a^2(M_t))$ coefficients from the vacuum polarization
contributions are: i) $-23.401$ (PT), ii) $-20.541$ (NA),
iii) $-19.647 $ ($\xiq=0$), iv) $-14.821$ ($\xiq=-3$), v) $-11.406$
(running coupling). We see that the vacuum polarization contributions are of
the same sign and roughly the same magnitude as the exact calculation.
However, the PT, NA, and $\xiq=0$ prescriptions overestimate the
$\Or(a^2(M_t))$ coefficient by 60.3\%, 40.7\%, and 34.6\%, respectively.
The fact that the NA does not provide an accurate determination is related to
the observation that when one applies the BLM rescaling \cite{BLM} in the
$\dr$ case, a rather large residual $\Or(a^2(M_t))$ term remains \cite{ps}.
On the other hand, the $\xiq=-3$ prescription approximates this coefficient
very accurately, within 1.6\%.
The leading $\xiq=-3$ calculation also gives an $\Or(a^2(M_t))$
coefficient $-8.476$ in $\mt/M_t$ and $+8.746$ in the inverse $M_t/\mt$
expansion. The exact results (in the $M_b=0$ limit) are $-9.125$ and $+10.903$
respectively, so that the $\xiq=-3$ approach underestimates the answer
 in these cases by 7.1\% and 22.3\%, respectively.
In summary,
we have seen that the $\xiq=-3$ BFG presents a number of interesting features:
i) within the BFG it generates the smallest constant accompanying
the logarithmic term, and is expected to provide the smallest asymptotic
coefficients in infrared-renormalon effects; ii) it coincides exactly with the
$\Or(\as)$ correction to the potential between two infinitely heavy quarks
\cite{hq};
iii) it approximates very well (within 1.6\%) the $\Or(a^2(M_t))$
coefficient in $\dq$; iv) it approximates quite well the corresponding
term in the $\mt/M_t$ expansion (within 7.1\%), although it is less precise
(22.3\%) in the inverse $M_t/\mt$ case.
In contrast with the other
approaches, the running coupling prescription underestimates
the $\Or(a^2(M_t))$ coefficient in $\dq$ by $\approx 22\%$ (although
it fares much better if the comparison is made with the contribution
to the exact result not involving the double triangle graph,
in which case it overestimates the answer by 9.1\%).

Assuming that the $v_n$ series (\Eq{dq} with $\rho_n\to v_n$)
provides also the dominant contributions for $n>1$, we can
employ it to estimate the higher order effects in $\dq$. We note that this
 expansion involves large higher order coefficients:
for example the coefficients of $a^3(M_t)$ and $a^4(M_t)$ are 81.3 and
1,435.2, respectively.
As an illustration, setting $M_t=175$GeV ($\as^{(5)}(175{\rm GeV})=0.10744
$ for $\as^{(5)}(\mz)=0.118$) and summing up to the previous to smallest
($n=6$) or smallest ($n=7$) terms, we find
$-\dq= 0.1219 $ or 0.1223 (in the evaluation we employ the exact
$\Or(a^2(M_t))$ coefficient $-14.594$ rather than $-14.821$). This is
close, albeit somewhat higher, than the value 0.1200 found in \Ref{ps}
by using the BLM optimization of $M_t/\mt$, together with other
considerations.
It is also somewhat larger than the values 0.1192, 0.1198, and 0.1197,
obtained \cite{ps} by applying directly the BLM \cite{BLM},
PMS \cite{PMS}, and FAC \cite{FAC} optimization procedures
to \Eq{12}, and larger by 0.0070 (0.0075) than the value 0.1149
from \Eq{12}.
As the coupling constant prescription gives the smallest renormalon
effects in Table 1, we have repeated the above calculation using the $r_n$
coefficients (retaining again the exact $\Or(a^2(M_t))$ term)
and find $-\dq= 0.1195$ ($n=5$) and 0.1198 ($n=6$),
which are even closer to the optimized values mentioned above.
It is worth noting that the 0.1149 value from \Eq{12}
is marginally consistent with the 5.3$\times10^{-3}$ error estimate around
 0.1200 given in \Ref{ps}. The renormalon calculations described above
favor, however, values of $\dq$ closer to the central value 0.1200 obtained
in that work.

In summary, the numerical evaluation
and theoretical asymptotic estimates of  leading vacuum polarization
contributions indicate the presence of a significant
contribution from the dominant infrared renormalons in $\dr$ when it
 is expressed in terms of the top pole mass, and its
absence when  the $\msbar$ running mass is employed.
The detailed results depend on the precise definition of the basic building
block, the gluon self-energy.
One of the vacuum polarization prescriptions, the $\xiq=-3$ BFG, leads to
an $\Or(a^2(M_t))$ coefficient that approximates very accurately
the result from the exact calculation.
Assuming that the infrared-renormalon effects evaluated with this prescription
also dominate the higher order contributions, one finds that they increase
the value of $\dq$ relative to \Eq{12} by an amount that is close to the
effects previously obtained by optimization considerations
\cite{ps,consid,yr}. A similar conclusion is reached if one simply employs
the running coupling $\as(\kappa)$ in the evaluation of the renormalon
effects.
\vskip 1cm
\subsection*{Aknowledgments}
The authors would like to thank A. Mueller, K. Philippides, M. Porrati,
and G. Weiglein for interesting discussions and communications.
This research was supported in part by the National Science Foundation under
grant No. PHY-9313781.

\newpage
\renewcommand{\arraystretch}{1.2}
\begin{table}
\[
\begin{array}{|r||r|r|r|r|r|r|}\hline
n & p_n \ \ \ & a_n\ \ \ & \ell_n\ \ \  & v_n \ \ \
& r_n\ \ \ & (v_n)_{as} \ \ \\  \hline\hline
0 & 1.07247 &1.07247  & 1.07247  & 1.07247  & 1.07247 &   \\  \hline
1 &2.28926 &2.00949  &1.92206 &1.44994  &1.11577 &1.40672 \\  \hline
2 &2.50023 & 1.93952  &1.77927 &1.03706 & 0.63733   &   1.02925 \\  \hline
3 &2.93381 & 2.35629  & 2.20476   &1.59249 &1.33432 & 1.58413    \\  \hline
4 &2.88349 & 2.19666  & 2.01084  & 1.18702  &0.73426  & 1.18785 \\  \hline
5 &3.00202 & 2.34266   & 2.17124   & 1.47723  & 1.17998  &  1.47672  \\  \hline
6 &2.96229 & 2.26905 & 2.08517   & 1.29537 & 0.88504  & 1.29543  \\  \hline
7 &2.99567 & 2.31705  & 2.13967  &   1.40676  &1.06946 & 1.40673 \\  \hline
8 &2.98016 &  2.29113 & 2.10957    &  1.34168 & 0.95921 &1.34169 \\  \hline
9 & 2.98995 & 2.30623 &  2.12695  &1.37910    &1.02336 &  1.37910\\  \hline
10 & 2.98490  &  2.29800& 2.11740 &1.35804   & 0.98684  & 1.35804  \\  \hline
\end{array}
\]
\caption{\sf The coefficients $\rho_n$ in Eq.\,(6). The $p_n$, $a_n$,
$\ell_n$, $v_n$, $r_n$ correspond to the pinch-technique, non-abelianization,
$\xiq=0$, $\xiq=-3$, and $\as(\kappa)$ prescriptions for $\Pi(-\kappa^2)$,
respectively. The last column gives the result of the
asymptotic estimates (Eqs.\,(12,13)) for the  $\xiq=-3$ case. The normalization
of Eq.(6) corresponds to $\rho_0=(1+\pi^2/3)/4=1.07247\ldots$
}
\end{table}
\begin{table}
\[
\begin{array}{|r||r|r|r|r|r|r|}\hline
n & \hat{p}_n \ \ \ & \hat{a}_n\ \ \ & \hat{\ell}_n\ \ \ & \hat{v}_n \ \ \ &
\hat{r}_n \ \ \ & (\hat{v}_n)_{as}\ \ \\ \hline \hline
0 & 0.07247 & 0.07247 &  0.07247   &  0.07247& 0.07247  &   \\  \hline
1 &  -0.49403 & -0.33359 &  -0.36905  & -0.20835& -0.37948 &-0.09436\\  \hline
2 & -0.34464 &  -0.26331 & -0.24266  & -0.19952 &  -0.12460 &-0.20072 \\ \hline
3 & 0.00058   & 0.06757 & 0.08718    &  0.17430   &  0.22649&0.16495\\  \hline
4 &-0.06959  &  -0.07918&   -0.08550  & -0.14297 & -0.20541& -0.14241\\  \hline
5 & 0.02159    &0.04066  & 0.04737  &   0.09628  &  0.15011 &0.09578 \\  \hline
6 &-0.01921  &-0.02726 & -0.03084  & -0.06154&  -0.09942  & -0.06148\\  \hline
7 & 0.00924   & 0.01521  & 0.01757 & 0.03712  &  0.06176 & 0.03710 \\  \hline
8 & -0.00576 & -0.00890  & -0.01023 & -0.02178 &  -0.03688 &-0.02178 \\  \hline
9 & 0.00308   & 0.00496 & 0.00574  & 0.01247   &   0.02141 &0.01247 \\  \hline
10 &-0.00174 & -0.00278 & -0.00321 & -0.00703 & -0.01219 &-0.00703  \\  \hline
\end{array}
\]
\caption{\sf Same as in Table 1, for the coefficients
$\hat{\rho}_n$ in Eq.\,(9). The normalization of Eq.\,(9) corresponds to
$\rhoc_0=\rho_0-1=0.07247\ldots$}
\end{table}

\end{document}